\documentclass[pra,twocolumn,amssymb]{revtex4}
\usepackage{graphicx}

\begin{document}

\title{Pairing between Atoms and Molecules in Boson-Fermion Resonant Mixture }
\author{Jian Zhang and Hui Zhai}\affiliation {Center for Advanced Study, Tsinghua
University, Beijing, 100084, P. R. China}
\date{\today}
\begin{abstract}
We consider a mixture of fermionic and bosonic atoms nearby
interspecies Feshbach resonances, which have been observed
recently in $^6$Li-$^{23}$Na mixture by MIT group, and in
$^{40}$K-$^{87}$Rb mixture by JILA group. We point out that the
fermion-boson bound state, namely the heteronuclear molecules,
will coexist with the fermionic atoms in a wide parameter region,
and the attraction between fermionic atoms and molecules will lead
to the formation of atom-molecule pairs. The pairing structure is
studied in detail, and, in particular, we highlight the possible
realization of the Fulde-Ferrel-Larkin-Ovchinnikov state in this
system.

\end{abstract}
\maketitle

{\it Introduction.} In the latest few years ultracold atomic gases
nearby Feshbach resonance have attracted considerable attention,
since Feshbach resonance can be utilized to achieve strongly
interacting quantum gases exhibiting universal properties, to
produce Bose-Einstein condensation of molecules from
atomic gases\cite{BEC}, and to study the physics of BEC-BCS
crossover\cite{BCS,BEC-BCS}. While most
present experimental and theoretical works focus on the resonance
between atoms of the same species, interspecies Feshbach
resonances were also observed recently in collisions between
fermionic $^6$Li and bosonic $^{23}$Na by MIT
group\cite{ketterle}, and in collisions between fermionic $^{40}$K
and bosonic $^{87}$Rb by JILA group\cite{Jin}. These interspecies
Feshbach resonances arise from the fermion-boson bound states,
which are called heteronuclear molecules. In this Rapid
Communication, we will show that the fermionic atoms and molecules
will coexist in a wide parameter region, and they will form
atom-molecule pairs and exhibit superfluidity, provided that the
life-time of molecules can be long enough to reach equilibrium.
Particularly, the system is a good candidate for realization of
the Fulde-Ferrell-Larkin-Ovichinnikov (FFLO) state, owing to the
mass difference between atom and molecule, as well as the tunable
molecular binding energy.

In the study of superconductor with mismatched Fermi surfaces of different
spins, FFLO state was proposed by Fulde and Ferrell\cite{FF},
and by Larkin and Ovchinnikov\cite{LO} independently. 
The distinct features of this state include (i) the
center-of-mass momentum of each pair is nonzero, and thus the
order parameter has a periodic space modulation; (ii) there are
unpaired normal fermions coexisting with paired superfluid
fermions. Stimulated by these exotic features, experimental
condensed matter physicists have paid a lot of efforts on
searching this state in various materials in the past decades, and
until recently some evidences of discovering this state in
$\text{CeCoIn}_{5}$ have been reported\cite{CeCoIn}. This state
has also been studied in the discussion of color superconductivity
and pulsars. These progresses have been systematically summarized
in a latest review article Ref.\cite{RMP}. Recently, the
realization of the FFLO state in the mixture of two-component
fermions has been suggested \cite{Combescot,Mizushima}, where the
Fermi surface mismatch is controlled by unequal population of
different components, and the experimental signatures of the FFLO
state in cold atoms experiments have also been
discussed\cite{Mizushima}. Besides, there are also some
other proposes for the pairing problem with mismatched Fermi surfaces, such as the breached-pair state\cite{BP} and pairing with deformed Fermi surface\cite{defomation}. Particularlly the energetic 
and dynamic stability of the breached-pair state have been discussed a lot in some recent literatures\cite{BP}.

{\it Setting up the model.} We consider a uniform mixture of
bosonic atoms (b.a.) and fermionic atoms (f.a.). Away from the
resonance, the properties of such a mixture has been extensively
discussed before\cite{BFmixture}. By introducing the operator
$\hat{\Phi}$ for the b.a., and the operator $\hat{\Psi}_{a}$ for
the f.a., this mixture can be described by
\begin{eqnarray}
\hat{H}_{\text{A}}=\int d^3{\bf r}\left[
\frac{\hbar^2}{2m_{1}}|\nabla\hat{\Psi}_{a}|^2+\frac{\hbar^2}{2m_{2}}|\nabla\hat{\Phi}|^2\right.\nonumber\\
\left.+g_{0}\hat{\Phi}^\dag\hat{\Phi}^\dag\hat{\Phi}\hat{\Phi}
+g_{1}\hat{\Psi}^\dag_{a}\hat{\Psi}_{a}\hat{\Phi}^\dag\hat{\Phi}\right],
\end{eqnarray}
where $g_{0}$ is the strength of the self-interaction between b.a,
$g_{1}$ is the strength of interaction between b.a. and f.a., and
$m_{1}$ and $m_{2}$ are the masses of f.a. and b.a. respectively.
Nearby the resonance, the process that converts two atoms into a
heteronuclear molecule (h.m.) should be incorporated. In the
spirit of the two-channel model\cite{Holland}, we introduce an independent fermionic
operator $\hat{\Psi}_{m}$ for this molecule, and
describe the resonance process as
\begin{equation}
\hat{H}_{\text{AM}}=\alpha\int d^3{\bf
r}\left(\hat{\Psi}^\dag_{m}\hat{\Phi}\hat{\Psi}_{a}+\hat{\Psi}^\dag_{a}
\hat{\Phi}^\dag\hat{\Psi}_{m}\right).
\end{equation}
And finally the molecules are described by
\begin{equation}
\hat{H}_{\text{M}}=\int d^3{\bf
r}\left[\frac{\hbar^2}{2m_{m}}|\nabla\hat{\Psi}_{m}|^2+\mu_{0}\hat{\Psi}^\dag_{m}\hat{\Psi}_{m}
+g_{2}\hat{\Psi}^\dag_{m}\hat{\Psi}_{m}\hat{\Phi}^\dag\hat{\Phi}\right].
\end{equation}
Here the mass $m_{m}$ of h.m., which equals $m_{1}+m_{2}$, will be much larger
than $m_{1}$ provided that $m_{2}$ is times larger than $m_{1}$. $g_{2}$ is
the interaction strength between h.m. and b.a., and the molecular
binding energy $\mu_{0}$ can be tuned by an external magnetic
field. Therefore the total Hamiltonian is given by $\hat
{H}=\hat{H}_{\text{A}}+\hat{H}_{\text{M}}+\hat{H}_{\text{AM}}$.
Since we are mostly interested in the behavior of the fermions in
a bosonic background, we consider the situation that the density
of bosons $n_{0}$ is much larger than the density of fermions $n$,
and therefore we can neglect the feedback on bosons safely. The
opposite situation where $n$ is comparable to $n_{0}$ has been
considered recently\cite{Powell}, and the
mean-field solution of the equal mixing case has also been
provided\cite{Stoof}.

\begin{figure}[tbp]
\begin{center}
\includegraphics[width=8.0cm]
{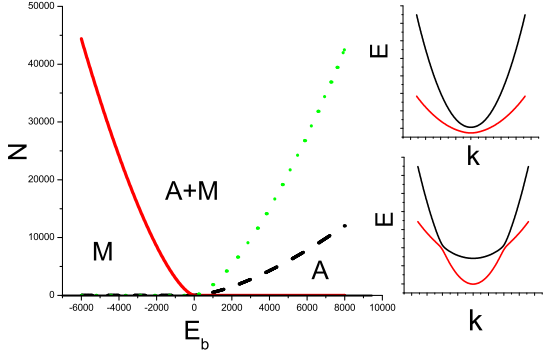} \caption{(color online) The mean-field
$N-E_{\text{b}}$ phase diagram. Below the red solid line is the
pure fermionic molecules phase; below the black dash line is the pure fermionic atoms
phase, and in the region between the solid line and dash line
atoms and molecules coexist. In the left-hand side of the dotted line, the Fermi
momentum of atoms $k_{\text{Fa}}$ is smaller than the Fermi
momentum of molecules $k_{\text{Fm}}$. The two insets schematically show the dispersion
relations $\epsilon_{\pm}({\bf k})$, with $E_{\text{b}}<0$ for the
upper one and $E_{\text{b}}>0$ for the lower one. The momentum $k$ is in the unit
of $k_{0}=L/(2\pi)$,
where $L$ is the length of the system, and the energy is in the unit of
$\hbar^2 k_{0}^2/(2m_{1})$. All the figures in this paper are plotted for
$^{40}$K-$^{87}$Rb mixture.
\label{amphase}}
\end{center}
\end{figure}

{\it The mean-filed solution.} At nearly zero temperature, the
bosonic atoms are condensed. Adopting the standard Bogoliubov
approximation, we expand the operator $\hat{\Phi}$ as
$\sqrt{n_{0}}+\sum_{{\bf k}}e^{i{\bf k}{\bf r}}\hat{\Phi}_{{\bf
k}}$. Neglecting some constants, the Hamiltonian at the mean-field
level turns out to be $\hat{H}_{0}=\sum_{{\bf k}}\epsilon_{a}({\bf
k})\hat{\Psi}^\dag_{a{\bf k}}\hat{\Psi}_{a{\bf
k}}+\epsilon_{m}({\bf k})\hat{\Psi}^\dag_{m{\bf
k}}\hat{\Psi}_{m{\bf
k}}+\alpha\sqrt{n_{0}}(\hat{\Psi}^\dag_{a{\bf
k}}\hat{\Psi}_{m{\bf k}}+\hat{\Psi}^\dag_{m{\bf
k}}\hat{\Psi}_{a{\bf k}}),$
where $\epsilon_{a}({\bf k})$ denotes $\hbar^2{{\bf k}}^2/(2m_{1})$, and
$\epsilon_{m}({\bf k})$ denotes $\hbar^2{{\bf k}}^2/(2m_{m})+E_{b}$. Here
the binding energy $\mu_{0}$ is normalized to $E_{\text{b}}$ by the mean-field
interactions, i.e. $E_{{\text{b}}}=\mu_{0}-g_{1}n_{0}+g_{2}n_{0}$. This
quadratic Hamiltonian can be easily solved, which yields
\begin{equation}
\epsilon_{\pm}=\frac{1}{2}\left[\epsilon_{a}({\bf
k})+\epsilon_{m}({\bf k})\pm \sqrt{(\epsilon_{a}({\bf
k})-\epsilon_{m}({\bf k}))^2+4\alpha^2n_{0}}\right],
\end{equation}
and they are shown in the insets of Fig. \ref{amphase}. Owing to
the large mass difference, $|\epsilon_{a}({\bf
k})-\epsilon_{m}({\bf k})|$ is usually very large unless ${\bf k}$
is very close to ${\bf k}_{\text{c}}$ where $\epsilon_{a}({\bf
k})$ crosses $\epsilon_{m}({\bf k})$, the dispersion relations
$\epsilon_{\pm}({\bf k})$ thus can be well approximated by
$\epsilon_{a(m)}({\bf k})$ when
$|k-k_{\text{c}}|\gg2\alpha\sqrt{n_{0}}m_{1}m_{m}/(m_{2}\hbar^2
k_{\text{c}})$, as can be seen from the Fig. \ref{amphase}. For a
narrow resonance, this approximation is valid in the momentum
region where the atom-molecule pairs are formed.

It is noticed that the total number of fermionic atoms, including
those in the scattering states and in the bound state, is
conserved during the resonance process, i.e.
$\hat{\Psi}^\dag_{a}\hat{\Psi}_{a}+\hat{\Psi}^\dag_{m}\hat{\Psi}_{m}=N$,
and these fermions fill the Fermi sphere with respect to
$\epsilon_{\pm}({\bf k})$ at zero temperature. Thus we obtain the
mean-field phase diagram for different $N$ and $E_{\text{b}}$
shown in Fig. \ref{amphase}, from which one can see that in a wide
range the atoms and molecules coexist in the system. This feature,
contrasting with the Feshbach resonance between the atoms of the
same species, arises from the fact that both atoms and molecules
in this system should obey the Pauli exclusive principle.
Moreover, the Fermi surfaces difference $\Delta k_{F}$, which
is defined as $k_{\text{Fm}}-k_{\text{Fa}}$, is given by $(\sqrt{2m_{\text{m}}
(\mu-E_{\text{b}})}-\sqrt{2m_{1}\mu})/\hbar$ for fixed chemical potential $\mu$,
and increases as the
decrease of $E_{\text{b}}$. This is of crucial importance for the
discussion of atom-molecule pairng below.

\begin{figure}[tbp]
\begin{center}
\includegraphics[width=7.8cm]
{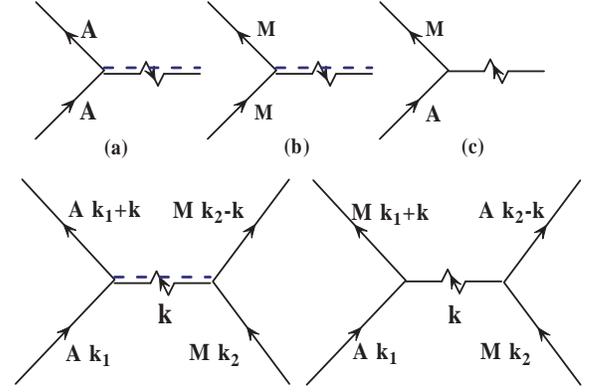} \caption{ Upper: A diagrammatical illustration of
the conventional scattering processes (a-b) and the Feshbach resonance process(c). Lower: The effective interaction
between fermions owing to exchange of density fluctuations. The straight solid lines denote the propagate of Fermions,
the twist solid lines denote the propagate of uncondensed Bosons, while the dotted lines denote the condensed Bosons.
\label{processes}}
\end{center}
\end{figure}

{\it Condensate fluctuation and the attractive interactions.} We now incorporate the coupling between fermions and the
condensate fluctuations illustrated in Fig. \ref{processes}, which includes (a) density-density interaction between f.a. and b.a.,
$g_{1}\sqrt{n_{0}}\hat{\Psi}^\dag_{a,{\bf
k}_{2}}\hat{\Psi}_{a,{\bf k}_{1}}(\hat{\Phi}^\dag_{{\bf
k}_{1}-{\bf k}_{2}}+\hat{\Phi}_{{\bf k}_{2}-{\bf k}_{1}})$; (b)
density-density interaction between h.m. and b.a.,
$g_{2}\sqrt{n_{0}}\hat{\Psi}^\dag_{m,{\bf
k}_{2}}\hat{\Psi}_{m,{\bf k}_{1}}(\hat{\Phi}^\dag_{{\bf
k}_{1}-{\bf k}_{2}}+\hat{\Phi}_{{\bf k}_{2}-{\bf k}_{1}})$, and
(c) the Feshbach resonance, $\alpha(\hat{\Psi}^\dag_{m,{\bf
k}_{2}}\hat{\Phi}_{{\bf k}_{2}-{\bf k}_{1}}\hat{\Psi}_{a,{\bf
k}_{1}}+\text{h.c.})$. Considering the second order processes that
exchange bosonic fluctuations between fermions, an effective
interaction between f.a. and h.m. will be induced\cite{Feynman},
which can be described by following effective Hamiltonian
\begin{eqnarray}
\hat{H}_{\text{int}}=\sum\limits_{{\bf k}_{1}{\bf k}_{2}{\bf
k}}\frac{1}{4gn_{0}+\frac{\hbar^2k^2}{2m_{2}}}\left(\alpha^2\hat{\Psi}^\dag_{m,{\bf
k}_{1}+{\bf k}}\hat{\Psi}^\dag_{a,{\bf k}_{2}-{\bf
k}}\hat{\Psi}_{a,{\bf k}_{1}}\hat{\Psi}_{m,{\bf
k}_{2}}\right.\nonumber\\
\left.-2g_{1}g_{2}n_{0}\hat{\Psi}^\dag_{m,{\bf k}_{2}-{\bf
k}}\hat{\Psi}^\dag_{a,{\bf k}_{1}+\bf {k}}\hat{\Psi}_{a,{\bf
k}_{1}}\hat{\Psi}_{m,{\bf k}_{2}}\right).
\end{eqnarray}
Given the typical values of $g_{1}$ and $g_{2}$, when the  density of bosonic atoms $n_{0}$ is as large as $10^{16}\text{cm}^{-3}$ for the width of resonance around $1\text{G}$, or when the width of the resonance is as narrow as $10\text{mG}$ for the density of bosonic atoms around $10^{14}\text{cm}^{-3}$, $2g_{1}g_{2}n_{0}$ will be larger than $\alpha^2$ and the interaction between atoms and molecules will be attractive. This attraction will lead to the
formation of atom-molecule pairs, and consequently the superfluid
order $\langle \hat{\Psi}_{a}\hat{\Psi}_{m}\rangle$.

Before proceeding, we should remark that not only the attraction
between f.a. and h.m., but also the attraction between f.a.( h.m.)
themselves will
be induced. Indeed these terms will result in an atom-atom (or
molecule-molecule) pairing order $\langle\sum_{{\bf
k}}\hat{\Psi}_{a,{\bf k}}\hat{\Psi}_{a,-{\bf k}}\rangle$. However,
In this work we focus on the pairng structure and superfluid properties of
atom-molecule paired state because the $p$-wave pairing between atoms (molecules) is usually
weak.

{\it Atom-molecule pairing.} The following discussion on the
pairing problem is based on the effective Hamiltonian for the
low-energy fermionic degrees of freedom, that is,
$\hat{H}_{\text{eff}}=\hat{H}_{0}+\hat{H}_{\text{int}}$. For the
low-energy scattering processes where the exchanged momentum $k$
is much smaller than $\sqrt{8gn_{0}m_{2}}/\hbar$, we can
approximate $\hat{H}_{\text{int}}$ by $-V\sum_{{\bf k_{1}}{\bf
k}_{2}{\bf k}}\hat{\Psi}^\dag_{m,{\bf
k_{2}-k}}\hat{\Psi}^\dag_{a,{\bf k_{1}+k}}\hat{\Psi}_{a,{\bf
k}_{1}}\hat{\Psi}_{m,{\bf k}_{2}}$, with $V$ denoting
$(2g_{1}g_{2}n_{0}-\alpha^2)/(4gn_{0})$. Following the idea of FF
and LO\cite{FF,LO,Takada}, we start with a generalized BCS order
parameter $\Delta_{q}=V\langle\sum_{{\bf k}}\Psi^\dag_{a,{\bf
k+q}}\Psi^\dag_{m,{\bf -k+q}}\rangle$, where ${\bf q}$ is
introduced as a variational parameter characterizing the
center-of-mass momentum of pairs. With the standard BCS mean-field
theory, we can obtain the excitation spectrums $E_{\pm}({\bf k})$
as
\begin{eqnarray}
&&E_{\pm}({\bf k})=\pm\frac{\epsilon_{a}({\bf
k+q})-\epsilon_{m}({\bf
-k+q})}{2}+\nonumber\\&&\sqrt{\left(\frac{\epsilon_{a}({\bf
k+q})+\epsilon_{m}({\bf -k+q})}{2}-\mu\right)^2+\Delta_{q}^2},
\end{eqnarray}
where $\mu$ is the chemical potential. Contrary to the
conventional BCS superconductor case, here $E_{\pm}({\bf k})$ are
not always positive for arbitrary $\Delta_{q}$, and the pairs will
break where the quasi-particle excitation energy is negative.
Therefore the wave function should be taken as
\begin{eqnarray}
|\varphi\rangle=\prod\limits_{\{{\bf k}|E_{+}({\bf
k})<0\}}\Psi^\dag_{a,{\bf k+q}}\prod\limits_{\{{\bf k}|E_{-}({\bf
k})<0\}}\Psi^\dag_{m,{\bf -k+q}}\nonumber\\\prod\limits_{\{{\bf
k}|E_{\pm}({\bf k})>0\}}\left(u_{{\bf k}}+v_{{\bf
k}}\Psi^\dag_{a,{\bf k+q}}\Psi^\dag_{m,{\bf
-k+q}}\right)|0\rangle.
\end{eqnarray}
The order parameter $\Delta_{q}$, as well as the depairing
regions, should be determined self-consistently from following gap
equation
\begin{equation}
1=V\sum\limits_{{\bf k}}\frac{1-\Theta(E_{+}({\bf
k}))-\Theta(E_{-}({\bf k}))}{\sqrt{\left(\frac{\epsilon_{a}({\bf
k+q})+\epsilon_{m}({\bf
-k+q})}{2}-\mu\right)^2+\Delta_{q}^2}},\label{gapequation}
\end{equation}
where $\Theta(x)$ is the step function, with $\Theta(x)=1$ for
$x<0$ and $\Theta(x)=0$ for $x>0$.

\begin{figure}[tbp]
\begin{center}
\includegraphics[width=4.3cm]
{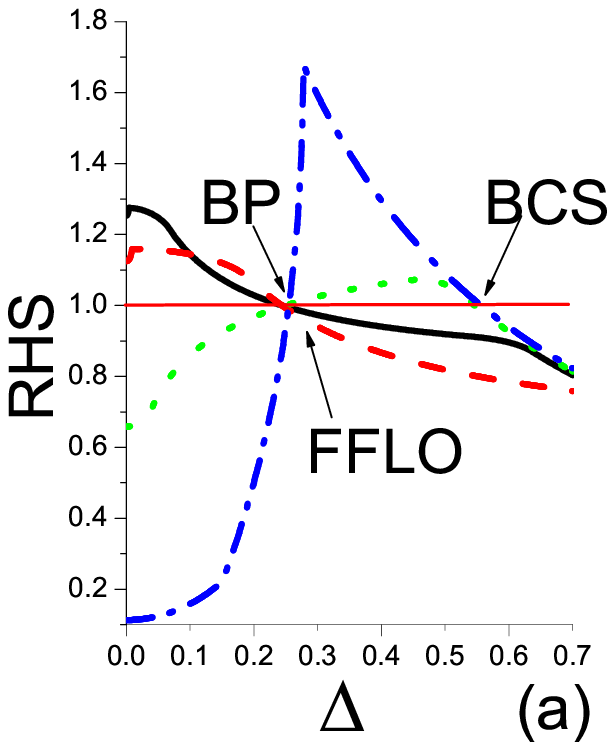}
\includegraphics[width=4.2cm]
{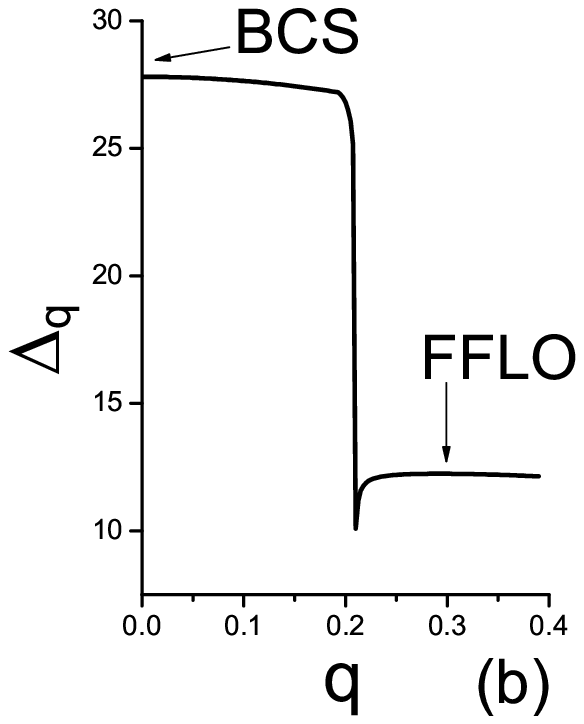} \caption{(color online) (a) The right-hand side (R. H.
S.) of the gap equation Eq.(\ref{gapequation}) as a function of
$\Delta$ with different values of $q$. $q=0$ for the dash-dotted
line, $q=0.175k_{\text{Fa}}$ for the dotted line, $q=0.255k_{\text{Fa}}$ for the solid line
and $q=0.35k_{\text{Fa}}$ for the dash line. (b) The solutions of the gap
equation Eq.(\ref{gapequation}), $\Delta_{q}$, as a function of
$q$. Both figures are plotted for the chemical potential $\mu$
fixed at $E_{\text{Fa}}$,  $\Delta k_{\text{F}}=0.41k_{\text{Fa}}$ and
$V=8\times 10^{-8}E_{\text{Fa}}$. The unit of $q$ is $k_{\text{Fa}}$, and the unit of 
$\Delta$, $\mu$ and $V$ is taken as $E_{\text{Fa}}=\hbar^2 k_{\text{Fa}}^2/(2m_{1})$ with $\hbar^2/(2m_{1})$ setting as unity.
 \label{pairing}}
\end{center}
\end{figure}

\begin{figure}[tbp]
\begin{center}
\includegraphics[width=4.2cm]
{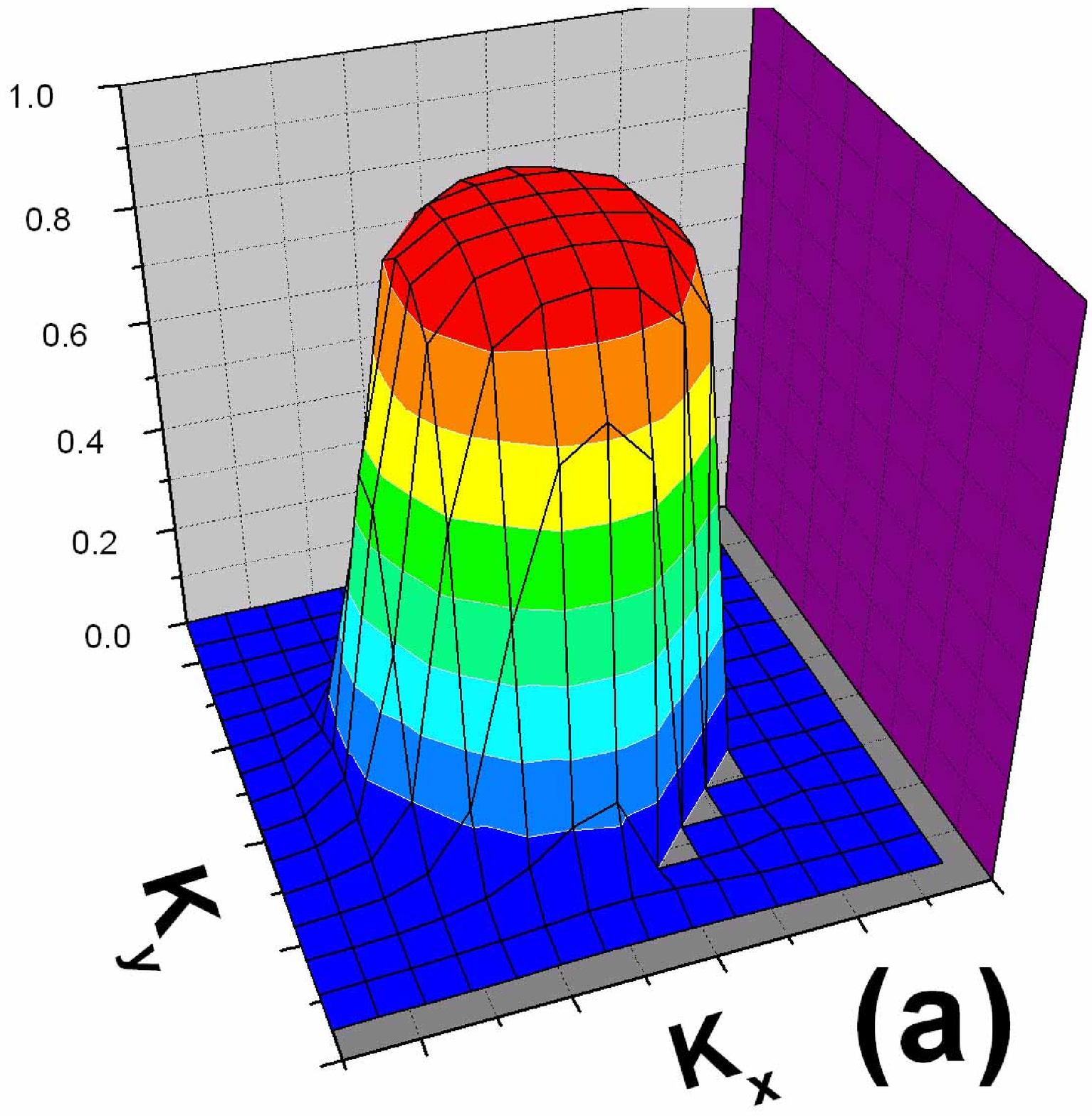}
\includegraphics[width=4.2cm]
{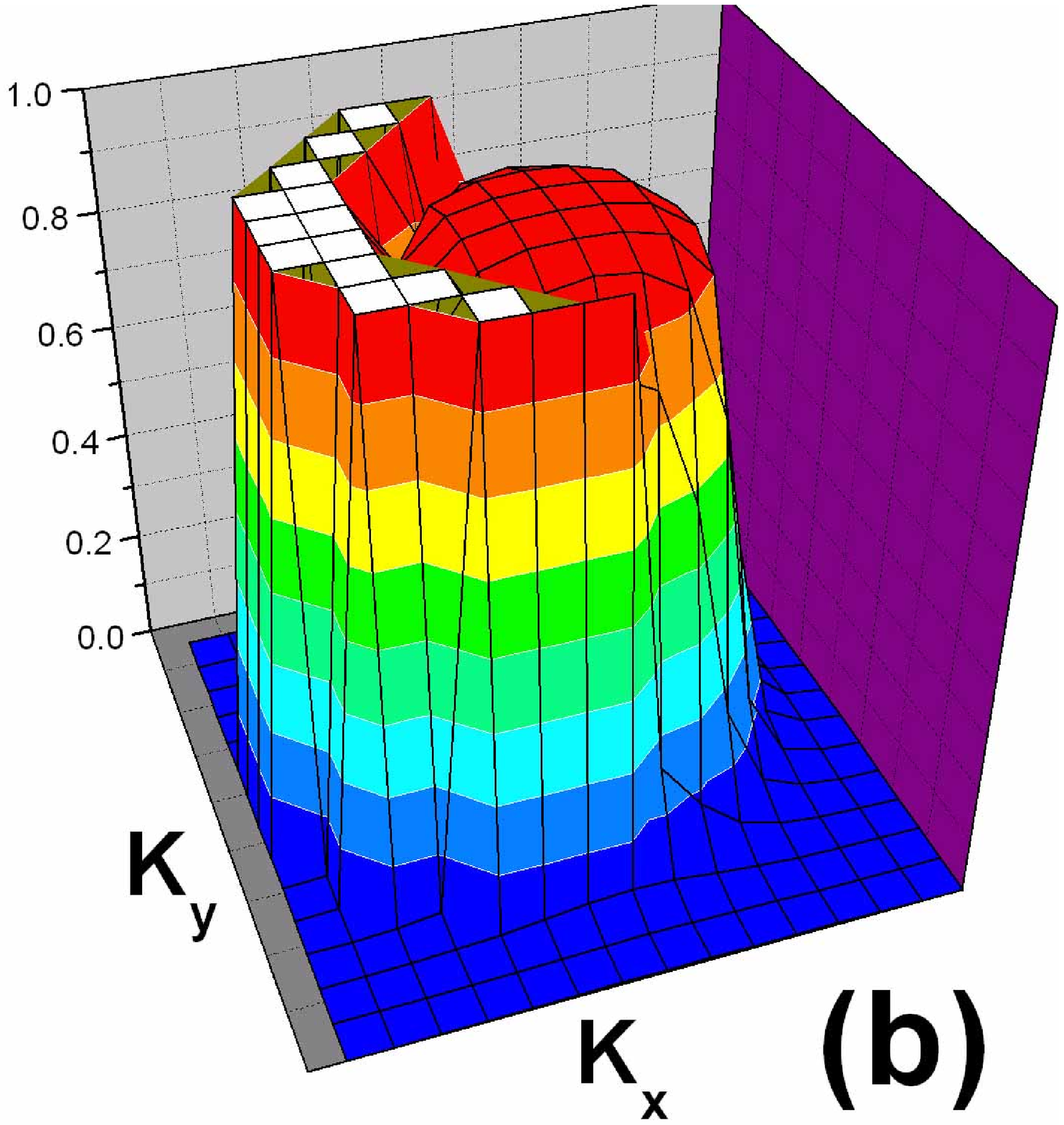} \caption{(color online) A illustration of the distribution of
atoms(a) and the distribution of molecules(b) in the FFLO state.
Here $\Delta k_{\text{F}}=0.41k_{\text{Fa}}$,
$q=0.255k_{\text{Fa}}$, and $V=8\times 10^{-8}E_{\text{Fa}}$.
\label{distribution}}
\end{center}
\end{figure}

In the Fig.\ref{pairing}(a) we plot the right-hand side of the gap
equation Eq.(\ref{gapequation}) as a function of $\Delta_{q}$ with
different values of $q$, from which we can know all the possible
solutions to the gap equation. We find that the equation for
$\Delta_{0}$ has two solutions. The one of large $\Delta$ is the
BCS solution, where Cooper pairs are formed around
$k_{\text{F}}=[(k^3_{\text{Fa}}+k^3_{\text{Fm}})/2]^{1/3}$ and the
excitation spectrums are all gapped; and another of small $\Delta$
corresponds to the breached-pair state. The energy of the breached-pair
state has been extensively studied before for similar model\cite{BP}
and will not be repeated here. The curves with small $q$, such as $q=0.175k_{\text{Fa}}$,
are similar to the curve with $q=0$. However, as $q$ increases the
gap equation has only one solution at small $\Delta$, which
corresponds to the FFLO states. The values of $\Delta_{q}$ are
obtained numerically for different $q$, and shown in
Fig.\ref{pairing}(b). The small $\Delta$ solutions, including the
breached-pair state and the FFLO states, are all characterized by
the existence of the deparing region, which can be found from
Fig.\ref{distribution} where the particle populations of a typical
FFLO state are shown.

\begin{figure}[tbp]
\begin{center}
\includegraphics[width=4.2cm]
{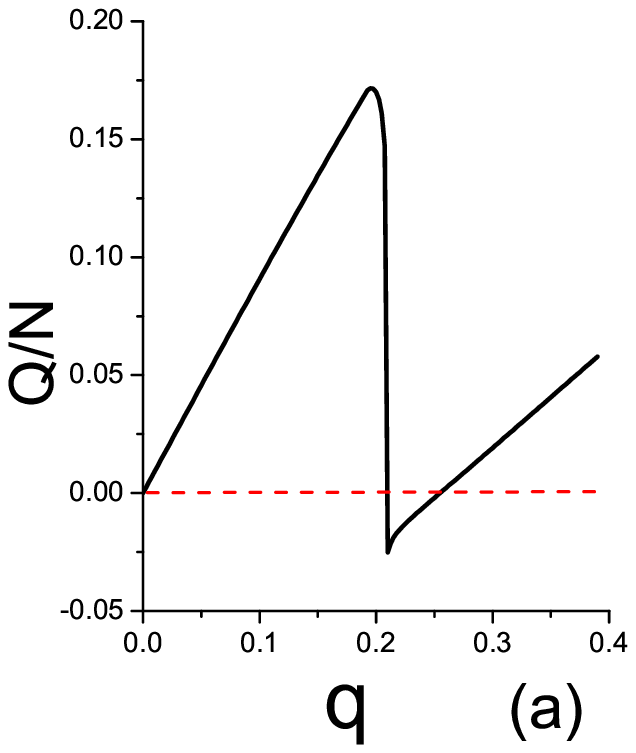}
\includegraphics[width=4.0cm]
{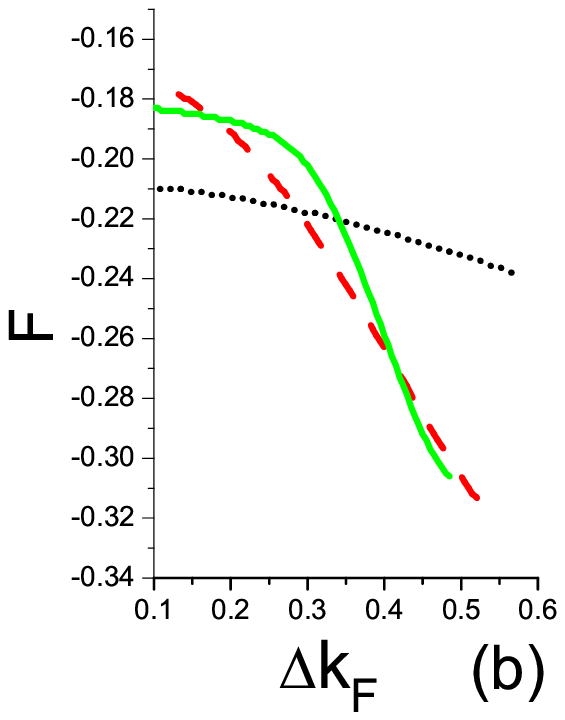} \caption{(color online)(a) The momentum per particle
$Q/N$ of the FFLO states for different values of $q$. $\Delta
k_{\text{F}}$ is fixed at $0.41k_{\text{Fa}}$. (b) The free energy per particle
$F=E/N-\mu$ as a function of $\Delta k_{\text{F}}$ for BCS state
(dotted line) and the FFLO states with $q=0.255k_{\text{Fa}}$(solid line) and
$q=0.30k_{\text{Fa}}$(dash line). We change $\Delta k_{\text{F}}$ by tuning
$E_{\text{b}}$. Both figures are plotted for $V=8\times 10^{-8}E_{\text{Fa}}$. The units of $q$, $Q$
and $\Delta k_{\text{F}}$ is $k_{\text{Fa}}$ and the unit of $F$ is $E_{\text{Fa}}$.
 \label{energy}}
\end{center}
\end{figure}

In the FFLO state, the center of Fermi sphere moves to ${\bf q}$.
The atoms pair with molecules at the small momentum side of the
Fermi sphere, while the pairs at the high-momentum side will
break. In this way, the FFLO states find a proper balance between
lowering the interaction energy and not costing too much kinetic
energy, and the nonzero momentum carried by the pairs can be
cancelled out by the depaired normal molecules, producing an
energetic stable FFLO state with the total momentum $Q$ vanishing.
We calculate the total momentum of the FFLO states for different
values of $q$ and find stable state exists for large $\Delta
k_{\text{F}}$, such as $\Delta k_{\text{F}}=0.41k_{\text{Fa}}$ as shown in Fig.
\ref{energy}(a), and in Fig. \ref{energy}(b) we also compare the
energy of two FFLO states with the BCS state, and find the FFLO
state will dominate over the BCS state with the increase of
$\Delta k_{\text{F}}$.

{\it Remark.}  Although we have demonstrated the existence of 
non-BCS type pairing state in certain region, the phase diagram of this system in the 
full parameter region is much richer. With the increase of $E_{\text{b}}$, the system will enter a pure 
atomic phase, the atom-molecule pairing will be replaced by the $p$-wave pairing between fermionic atoms.
Provided that the electric moments of these heteronuclear molecules 
are polarized by an external electric field, the direct dipole-dipole
interactions between molecules can not be neglected, and it will lead to the pairing between molecules\cite{dipole}. The 
competition between these different superfluid orders will manifest itself in quantum phase transitions, and more
fruitful physics is expected from the interplay of these phases, which will
be a subject for further inverstigations.

The authors would like to thank Professor C. N. Yang for
encouragement and valuable support, and we thank L. Chang, T. L.
Ho, F. Zhou, Z. Z. Chen, R. L\"u and M. Zhang for helpful
discussions, and we would like to thank Q. Zhou for
critical reading the manuscript. This work is supported by NSF
China (No. 10247002 and 10404015).

\end{document}